\documentclass[reprint, amsmath, amssymb, superscriptaddress]{revtex4-2}

\usepackage{graphicx}
\graphicspath{{./plots/}}
\usepackage{natbib}
\usepackage{amsmath}
\usepackage{dcolumn}
\usepackage{bm}
\usepackage[colorlinks=true,citecolor=blue,linkcolor=blue,urlcolor=blue]
{hyperref}

\begin{document}
	
	\title{Geodesic Motion and Particle Confinements in Cylindrical Wormhole Spacetime: Exploring Closed Timelike Curves.}
	
	
	\author{Dhritimalya Roy}
	\email{rdhritimalya@gmail.com}
	\affiliation{Department of Physics, Jadavpur University, Kolkata-700032, INDIA}
		\author{Ayanendu Dutta}
	\email{ayanendudutta@gmail.com}
	
	\affiliation{Department of Physics, Jadavpur University, Kolkata-700032, INDIA}
	
	\author{Subenoy Chakraborty}
	\email{schakraborty.math@gmail.com}
	\affiliation{Department of Mathematics, Jadavpur University, Kolkata-700032, INDIA}

	
\begin{abstract}
	In this study, the geodesic motion of a test particle along with its confinement is  investigated within Cylindrically Symmetric Wormhole spacetime admitting to Closed Timelike Curves. The confinement of particles with or without angular momentum is also investigated.  It is found that  particles with positive angular momentum that co-rotates with the spacetime can only pass through the causality violating region. Particles with only non-zero angular momentum are present in the vicinity of the Closed Timelike Curves.\end{abstract}
\maketitle

\section{Introduction}\label{introduction}
Wormholes are interconnecting ``bridges" or ``tunnels" which can connect two different universe or two points (at infinity) within the same universe. Wormholes are an interesting outcome of the solution of Einstein's GTR and was first introduced by Einstein and Rosen \cite{ER}, and termed it as The Einstein-Rosen Bridge. Traversability of Wormholes was first introduced by Morris and Throne \cite{MT}, since then various aspects of traversability are being studied at great lengths. Existence of Wormholes with respect to the validation of energy conditions have been studied \cite{MT,Visser,CNHC} and it is found that the stress energy tensor of matter violates Null Energy Condition (NEC) near the throat. These conditions are deduced under the assumption that, the throat is of 2-D geometry having a finite area, for a static case. Other definition of wormhole throat like for dynamic wormholes a for generalised notion of a wormhole throat was studied and  put forward, Readers may refer to these articles for better understanding \cite{hocheberg_visser,Hayward}.

The Entrance of a Wormhole (owing to its description in the frame work of spherical symmetry) generally implied, as seen from outside to be a local object like black holes and stars. But our universe may contain structures which are extended along certain directions, like cosmic strings. So to describe the simplest string like structure or configuration we use cylindrical symmetry. Wormholes in cylindrical symmetry along with its properties are studied by authors \cite{bronnikov1,bronnikov2,bronnikov3,bronnikov4,Forghani, Eiroa}, one may refer to the articles for elaborate reading.

Particle motion in the wormhole spacetime is an interesting topic of investigation, especially in traversable wormhole. There are numerous studies done by various authors that have investigated the motion of particles around the Wormhole geometry \cite{Cataldo,Olmo,CC,Ellis}. 
Motion of particles within or in the vicinity of Closed Timelike Curves is not the same as the motion of particles in general within the Wormhole spacetime.

Motivated from the above studies, in the present work we analyse the geodesic motion of a test particle in Cylindrical Wormhole Spacetime. Here, in this article we will investigate in details the motion of circular orbits of  test particles along the radial coordinate on the constant z-plane of the wormhole spacetime. We consider the orbits in constant Z-plane because they are the most effective way to measure the effects of Closed Timelike Curves admitted in this spacetime. We begin with a brief introduction of the  Wormhole metric and the Field Equation followed by the the Geodesic formulation along with the profile of Effective potential and its effects on the trajectories and velocities.

\section{Cylindrical Wormholes}
\subsection{The wormhole metric: A brief review}

The simplest form of the cylindrically  symmetric Wormhole metric is given as \cite{bronnikov1, bronnikov3},
\begin{equation}\label{eqn1}
	ds^2 = Adx^2+Cdz^2+Bd\phi^2+Edtd\phi-Ddt^2.
\end{equation}

We can rewrite (\ref{eqn1}) as,
\begin{equation}\label{eqn2}
ds^2= -D[dt-\frac{E}{D}d\phi]^2+ Adx^2+Cdz^2+Bd\phi^2 ,
\end{equation}

where, 
\begin{equation}\label{eqn3}
D=e^{2\gamma},A= e^{2\alpha}, C= e^{2\mu}, r(x)= e^{\beta}, r^2= \frac{\delta}{D}, \delta=BD+E^2.
 \end{equation}
 
 Hence, the metric becomes

\begin{eqnarray}\label{eqn4}
\nonumber ds^2= -e^{2\gamma(x)}\left[dt-\frac{E}{e^{2\gamma(x)}}d\phi \right]^2+ e^{2\alpha(x)}dx^2 \\+e^{2\mu(x)}dz^2+e^{2\beta(x)}d\phi^2 .
\end{eqnarray}

where, $E(x)$ is the rotation term and putting $E(x)=0$, simply gives us the static case.

Here, 
\begin{equation}\label{eqn14}
e^{2\gamma(x)} = Q^2 (x_{0}^2 - x^2) , 
\end{equation}
\begin{equation}\label{eqn15}
r^2 = e^{2\beta} = \left(\frac{r_{0}^2}{Q^2(x_{0}^2 - x^2)}\right),
\end{equation}
\begin{equation}\label{eqn16}
x_{0} = \frac{\omega_{0}}{(\chi \rho_{0} r_{0})} ,~~~ ~~
Q^2 = \chi \rho_{0} r_{0}^2.
\end{equation}

Here, $Q$ and $x_{0} $ are dimensionless constants and $r $ and $e^{\alpha}$ have the dimensions of length.

\begin{equation}\label{eqn17}
e^{2\mu} = e^{2mx}(x_{0}-x)^{(1-\frac {x}{x_{0}})} (x_{0}+x)^{(1+\frac {x}{x_{0}})}  ,~~
\end{equation}
where m is a constant.

And lastly , 
\begin{equation}\label{eqn18}
E= \frac{r_{0}(x_{0}^2-x^2)}{2x_{0}^2} \left(\frac{2x_{0}x}{(x_{0}^2-x^2)} + \ln{\frac{x_{0}+x}{x_{0} - x}} + E_{0} \right) .
\end{equation}
where $E_{0}$ is a constant.

In terms of harmonic radial co-ordinate $x$,
\begin{equation}\label{eqn19}
\alpha= \beta + \gamma + \mu.
\end{equation}

Equation (\ref{eqn14}) , (\ref{eqn15}), (\ref{eqn16}) , (\ref{eqn17}), (\ref{eqn18}) are obtained by simplifying and integrating the diagonal components of the Einstein equation. Readers may refer to \cite{bronnikov4} for detailed calculation.

\subsection{Field Equations: A brief review}

For the metric given by (\ref{eqn4}) describes a wormhole if \cite{ bronnikov3, bronnikov4} $
i) r(x) = e^{\beta(x)} $ has regular minimum and is large or finite away from this minimum. It is called the r-throat, or
ii)for the area function the same condition holds i.e $ a(x)= e^{\mu+\beta}$, also has the regular minimum, called the a-throat.
The term $``-E(x)"$ is the spacetime rotation term and is characterised by $\omega(x)$ i.e the angular velocity:
\begin{equation}\label{eqn5}
\omega= \frac{1}{2} \left(E e^{-2\gamma} \right)' e^{\gamma - \beta - \alpha}.
\end{equation}

The field equation along with the Ricci tensor components in the gauge $\alpha = \mu$  is given by \cite{bronnikov4} (prime denotes $\frac{d}{du}$) :-
\begin{equation}\label{eqn6}
R^{1}_{1} = -e^{-2\mu} \left[\beta'' + \gamma'' + \mu'' + \beta'^2 + \gamma'^2 + \mu'(\beta' + \gamma') \right] + 2\omega^2 ;
\end{equation}
\begin{equation}\label{eqn7}
R^{2}_{2} = -e^{-2\mu}[\mu'  + \mu'(\beta' + \gamma')]  ;
\end{equation}
\begin{equation}\label{eqn8}
\sqrt g R^{3}_{3} = -[\beta' e^{\beta+\gamma}  - E\omega e^{\mu}]' ;
\end{equation}
\begin{equation}\label{eqn9}
\sqrt g R^{4}_{4} = - \left(\omega e^{2\gamma + \mu} \right)' ;
\end{equation}
\begin{equation}\label{eqn10}
R^{4}_{3} = 0 ;
\end{equation}
The diagonal part of the Ricci tensor splits into two parts:-
i) the static part $ {}_{s}R_{\mu}^{\nu}$, and 
ii) for the rotational part $ {}_{\omega}R_{\mu}^{\nu}$, where,
\begin{equation}\label{eqn11}
{}_{\omega}R_{\mu}^{\nu} = \omega^2 dia(2,0,2,-2),
coordinate ~order ~(x,z,\phi,t).
\end{equation}

Similarly $G_{\mu}^{\nu}$ also splits in two , i.e 
\begin{equation}\label{eqn12}
G_{\mu}^{\nu} = {}_{s}G_{\mu}^{\nu} + {}_{\omega}G_{\mu}^{\nu} ,
~~ where, ~~
{}_{\omega}G_{\mu}^{\nu} = \omega^2 dia(1,-1,0,-3).
\end{equation}

The Einstein's equation can be written as :
\begin{equation}\label{eqn13}
G_{\mu}^{\nu} = -8 \pi G T_{\mu}^{\nu},
\end{equation}
and the stress energy tensor is given by :
\begin{equation}
R_{\mu}^{\nu} = -8 \pi G \tilde{T}_{\mu}^{\nu} ,
~~
where ,~~
\tilde{T}_{\mu}^{\nu}=  T_{\mu}^{\nu} - \frac{1}{2} \delta_{\mu}^{\nu} T_{\alpha}^{\alpha}.
\end{equation}
One may refer to \cite{ bronnikov3, bronnikov4} for the detailed study of the solution of EFE for cylindrical wormholes.

\section{Geodesic Formulation}\label{III}

For the formulation of the geodesic equation we use the cylindrically symmetric metric given by (\ref{eqn4}) , i.e:
\begin{eqnarray}
\nonumber ds^2= -e^{2\gamma(x)}\left[dt-\frac{E}{e^{2\gamma(x)}}d\phi \right]^2+ e^{2\alpha(x)}dx^2 \\+e^{2\mu(x)}dz^2+e^{2\beta(x)}d\phi^2 .
\end{eqnarray}

The geodesics of the spacetime can be calculated using the Euler-Lagrange equation of motion.
From the metric given by equation, (\ref{eqn4}), the Lagrangian becomes:

\begin{eqnarray}
\nonumber \mathcal{L} =  \frac{1}{2} (-e^{2\gamma(x)} \dot{t}^2 + \left[e^{2\beta(x)} - e^{-2\gamma(x)} E^2(x)\right] \dot{\phi}^2\\ + 2 E(x) \dot{\phi} \dot{t}   +e^{2\alpha(x)} \dot{x}^2 + e^{2\mu(x)} \dot{z}^2).
 \end{eqnarray}
 
  From the Lagrangian equation of motion the geodesic equations are obtained as follows:
  
 \begin{equation}\label{eqn20}
\dot t = - \frac{-A e^{2\beta(x) }e^{2\gamma(x)} - p_\phi e^{2\gamma(x)} E + AE^2}{e^{2\beta(x)} (e^{2\gamma(x)})^2} ,\end{equation}
 
 \begin{equation}\label{eqn21}
\dot \phi =   - \frac{ - p_\phi e^{2\gamma(x)}  + AE}{e^{2\beta(x)} e^{2\gamma(x)}},
\end{equation}
\begin{equation}\label{eqn22}
p_{z} = e^{2\mu(x)} \dot{z},
\end{equation}
and,

\begin{widetext}
\begin{equation}\label{eqn23}
\dot{x}^2 = \frac{- \epsilon+ e^{2\gamma(x)} \dot{t}^2 - \left[e^{2\beta(x)} - e^{-2\gamma(x)} E^2(x)\right] \dot{\phi}^2 \\ - 2 E(x) \dot{\phi} \dot{t} - e^{2\mu(x)} \dot{z}^2} {e^{2\alpha(x)}}.
\end{equation}
\end{widetext}

Here $A$ , $p_{\phi}$ , $p_{z}$ are integration constants and are termed as Energy of the particle, angular momentum and momentum along the $z$-direction respectively. $\epsilon$, here can take the values 1,0,-1  for timelike, null and spacelike signature of metric respectively.

In equation(\ref{eqn21}), it can be seen that it depends on the angular momentum of the test particles. If the angular momentum of the test particle is considered to be zero i.e $p_{\phi} = 0$, equation becomes:
\begin{equation}\label{SD}
\dot \phi =   - \frac{AE}{e^{2\beta(x)} e^{2\gamma(x)}}.
\end{equation}

The above expression may be termed as the equation of ``\textit{Space-time dragging}". It is proportional to the rotation term of the spacetime. As a result of the above equation , it can be deduced that the zero angular momentum particles also co-rotates with the rotating spacetime. Thus there are two types of co-rotating particles with $p_{\phi} = 0$, and $p_{\phi} > 0$. Particles with negative angular momentum i.e $p_{\phi} < 0$ are considered counter rotating particles.

Bronnikov \textit{et. al} in their paper \cite{bronnikov4} mentioned that their is a causality violation in the form of CTC in the general description of the spacetime. It emerges from the solution of the described spacetime.  It can be seen that if the $g_{\phi\phi}$ of the metric given by equation (\ref{eqn4}), is less than zero, it contains a closed timelike curve.
From the inspection of the $g_{\phi\phi}$, which is given by:
\begin{equation}\label{eqn24}
g_{\phi\phi} = \frac{r_{0}}{(x_{0}^2 - x^2)} \left[-1+(y + \frac{1-y^2}{2} \ln{\frac{(1+y)}{(1-y)}})^2 \right].
\end{equation}
shows, that the $g_{\phi\phi}$ component becomes negative and there occurs a CTC at $|y| > 0.564$ and there also occurs singularity at $|y| = \pm{1}$ , here $y=x/x_{0}$.
 
 Hence, it is of interest to investigate the trajectories of the particles in the vicinity of the CTC and also look for their confinements.

\section{Particle motion and Confinements}
Geodesic motion of a particle is allowed only when it satisfies the condition \cite{FW},
 \begin{equation}
\left(\frac{dx}{d\phi} \right)^2 \ge 0.
\end{equation}

The co-rotating particles satisfy the above condition while the counter rotating particles violates the condition for the positive radial distance. Hence from here we will only consider the co-rotating particles for our investigation. The motion of the test particles in the vicinity of Closed Timelike Curves are obtained by solving the $ (dt/dx )$ equation for null and time-like particles numerically,  as it is quite difficult to obtain analytic solutions for the first order geodesic equations. 

The solutions for the Radial Null Geodesic and Radial Timelike Geodesic is plotted along the radial distance (maintaining co-rotation of particles with the spacetime) to get the spacetime diagram to study their confinements.

\subsection{Radial Null Geodesic}
The radial null geodesic equation is obtained by simply using $\epsilon = 0$ in equation(\ref{eqn23}).
Hence, we get,

\begin{widetext}
\begin{equation}\label{eqn25}
\dot{x}^2 = \frac{ e^{2\gamma(x)} \dot{t}^2 - \left[e^{2\beta(x)} - e^{-2\gamma(x)} E^2(x)\right] \dot{\phi}^2 \\ - 2 E(x) \dot{\phi} \dot{t} - e^{2\mu(x)} \dot{z}^2} {e^{2\alpha(x)}}.
\end{equation}
\end{widetext}

The trajectories of the null particles or photons are obtained by solving the equation,

\begin{equation}
\frac{dt}{dx} = \frac{\dot {t}}{\dot{x}},
\end{equation}

From equations (\ref{eqn20}), (\ref{eqn21}), and (\ref{eqn25})  we get,
\begin{equation}\label{eqn26}
\frac{\dot {t}}{\dot{x}} = \frac{\left(-\frac{-A e^{2\beta}e^{2\gamma} - p_(\phi)e^{2\gamma} + E + A E^2}{e^{2\beta}(e^{2\gamma})^2}\right)}{\left(\frac{\frac{2Ap_{\phi}E}{e^{2\beta}} -( \frac{p_{\phi}^2 e^{2\gamma}}{e^{2\beta}} + A^2(-1 + \frac{E^2}{e^{2\gamma}e^{2\beta}}) + \frac{p_{z}^2 e^{2\gamma}}{e^{2\mu}})}{e^{2\beta} e^{2\mu} (e^{2\gamma})^2}\right)^{\frac{1}{2}}}.
\end{equation}

Here,  $\dot{x}^2$ has two roots, and for $\dot{x}$ has to be real for the solution to make sense.This can be readily achieved by carefully choosing the values of the constants. The values so chosen must not alter any conditions that allows the formation of the wormhole and the CTC.
The values of m and $E_{0}$ need to be Zero to maintain the signature $g_{\phi\phi} < 0$ and also the value of $r_{0}< 1.041$ for causality violation to occur. The value of A needs to be greater than 1, otherwise it may encounter $\frac{1}{0}$ form during computation.

It is of significance to note here that the study of both the co-rotating type of particles i.e with zero angular momentum and non zero angular momentum is  important, because of the presence of the space-time dragging term, as the zero angular momentum particles also co-rotates with the rotating spacetime. Consequently, it is of interest to see the confinements of both type of test particles. We here only consider the movements of the particles in constant z-plane, ergo '$p_{z} = 0$'.

\begin{itemize}
  \item Zero Angular Momentum Particles ($p_{\phi} = 0$) 
  
  Equation (\ref{eqn26}) for \textit{$p_{\phi} = 0$} becomes:

  \begin{equation}\label{eqn28}
\frac{\dot {t}}{\dot{x}} = \frac{\left(-\frac{-A e^{2\beta}e^{2\gamma}  + A E^2}{e^{2\beta}(e^{2\gamma})^2}\right)}{\left(\frac{ - A^2(-1 + \frac{E^2}{e^{2\gamma}e^{2\beta}})}{e^{2\beta} e^{2\mu} (e^{2\gamma})^2}\right)^{\frac{1}{2}}},
\end{equation}

   \item Non-Zero Angular Momentum Particles ($p_{\phi} \neq 0$) 
  
  For \textit{$p_{\phi} \neq 0$} equation (\ref{eqn26}) remains same, i.e:
  
  \begin{eqnarray}
\frac{\dot {t}}{\dot{x}} = \frac{\left(-\frac{-A e^{2\beta}e^{2\gamma} - p_(\phi)e^{2\gamma} + E + A E^2}{e^{2\beta}(e^{2\gamma})^2}\right)}{\left(\frac{\frac{2Ap_{\phi}E}{e^{2\beta}} -( \frac{p_{\phi}^2 e^{2\gamma}}{e^{2\beta}} + A^2(-1 + \frac{E^2}{e^{2\gamma}e^{2\beta}}) + \frac{p_{z}^2 e^{2\gamma}}{e^{2\mu}})}{e^{2\beta} e^{2\mu} (e^{2\gamma})^2}\right)^{\frac{1}{2}}}. \nonumber
\end{eqnarray}
 \end{itemize}

 Now using values from equations  (\ref{eqn14}) , (\ref{eqn15}), (\ref{eqn16}) , (\ref{eqn17}), (\ref{eqn18}), in equations  (\ref{eqn26} and \ref{eqn28}) , and after numerically solving the equation, and maintaining $\dot{x}$ remains real. The space-time diagram for photons with $p_{\phi} = 0$ and $p_{\phi} = 1$ is obtained.

\begin{figure}[h]
	\centerline{\includegraphics[scale=.45]{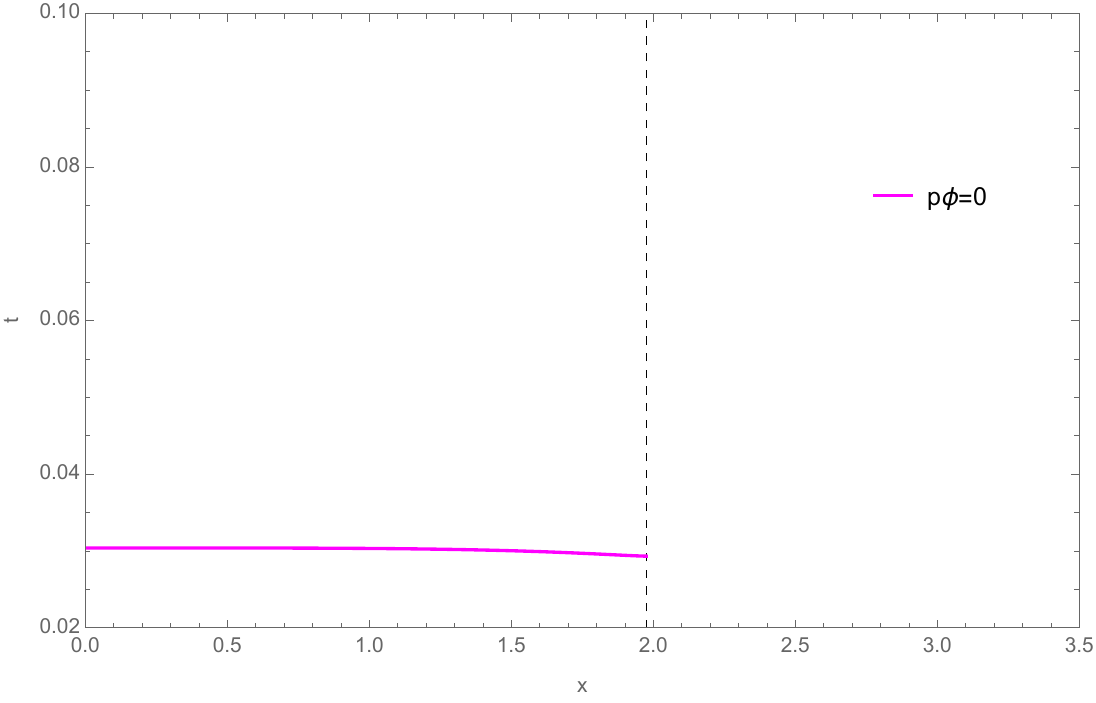}}
	\caption{Space-time diagram for radial null geodesic with $p_{\phi}= 0$.
	Here the values used are: $x_{0} = 3.5, \rho_{0}= .4, r_{0} = 1, \chi=.5, m=E_{0}=0, A = 3$. Particles are confined within $x= 1.974$.}
	\label{plot1}
\end{figure}

 \begin{figure}[h]
	\centerline{\includegraphics[scale=.7]{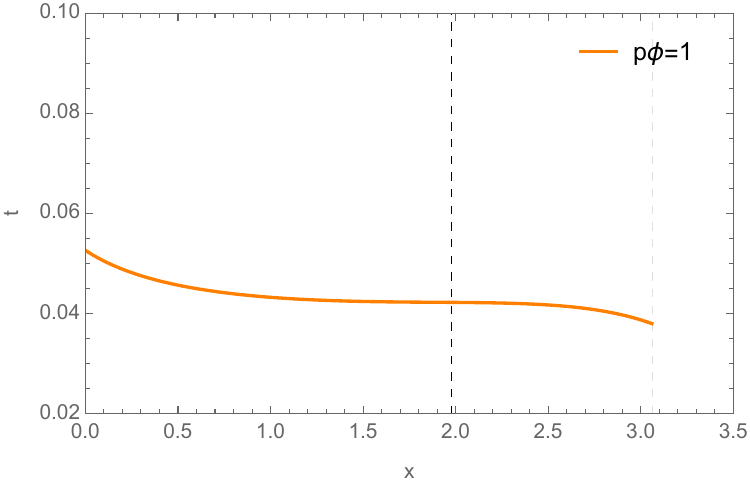}}
	\caption{Space-time diagram for radial null geodesic with $p_{\phi} = 1$.
	Here the values used are: $x_{0} = 3.5, \rho_{0}= .4, r_{0} = 1, \chi=.5, m=E_{0}=0, A = 3$. Particles are confined within $x= 3.059$.}

	\label{plot2}
\end{figure}

Figures (\ref{plot1}) and (\ref{plot2}) illustrates the spacetime diagram for particles with Zero AM and particles with Non-zero AM. It is apparent form the figures that both the test particles have different confinements (denoted by the dashed lines). The Zero AM particles are confined with in the region $x = 1.974$, which also happens to be the boundary of the CTC (for specific chosen values of constants). Contrastingly, particles having Non-zero AM are confined with in $x = 3.059$, breaching the CTC boundary. The results are somewhat similar to the paper \cite{AD} which studied motion of particles in van Stockum space-time.
This phenomenon can be interpreted as, the result of the Total Effective AM (i.e. AM of the spacetime + AM of the particle ) which allows the particle to breach the CTC boundary, and traverse in the CTC region.Velocity Profile of the above said particles for the same constant values also supports the above finding.

\begin{figure}[h]
	\centerline{\includegraphics[scale=.7]{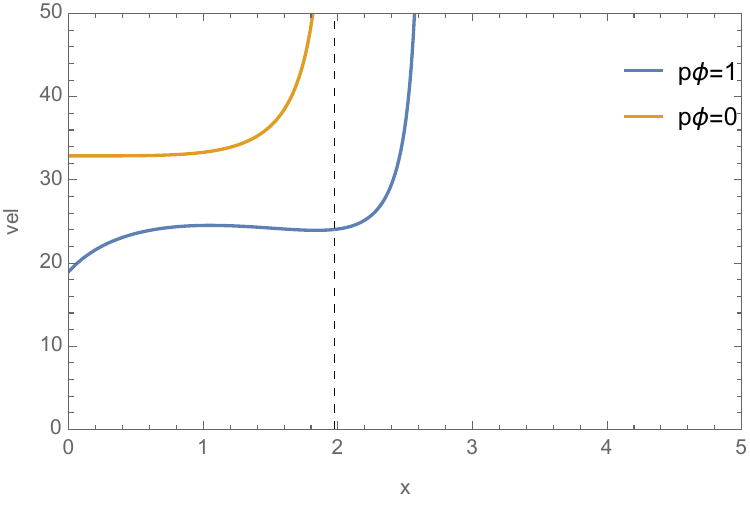}}
	\caption{Velocity profile graph for Null particles with $p_{\phi} = 0$, and $p_{\phi} = 1$. Particles with AM can only cross the CTC boundary denoted by the dashed lines.}

	\label{plot3}
\end{figure}

\subsection{Radial time like geodesic}

We can obtain the radial timelike geodesic from equation (\ref{eqn23}) by choosing $ \epsilon =1$, i.e

\begin{widetext}
\begin{equation}\label{eqn29}
\dot{x}^2 = \frac{-1+ e^{2\gamma(x)} \dot{t}^2 - \left[e^{2\beta(x)} - e^{-2\gamma(x)} E^2(x)\right] \dot{\phi}^2 \\ - 2 E(x) \dot{\phi} \dot{t} - e^{2\mu(x)} \dot{z}^2} {e^{2\alpha(x)}}.
\end{equation}
\end{widetext}

Again proceeding as before to get the spacetime diagram we use

\begin{equation}\label{eqn30}
\frac{\dot {t}}{\dot{x}} = \frac{\left(-\frac{-A e^{2\beta}e^{2\gamma} - p_(\phi)e^{2\gamma} + E + A E^2}{e^{2\beta}(e^{2\gamma})^2}\right)}{\left(\frac{\frac{2Ap_{\phi}E}{e^{2\beta}} -( \frac{p_{\phi}^2 e^{2\gamma}}{e^{2\beta}} + A^2(-1 + \frac{E^2}{e^{2\gamma}e^{2\beta}}) + \frac{p_{z}^2 e^{2\gamma}}{e^{2\mu}} + e^{2\gamma})}{e^{2\beta} e^{2\mu} (e^{2\gamma})^2}\right)^{\frac{1}{2}}}.
\end{equation}

As for the previously discussed reason in this subsection above, for the radial time-like particle also we investigate the confinements of the particles without AM and also with AM.
 
\begin{itemize}
  \item Zero Angular Momentum Particles ($p_{\phi} = 0$) 
  
  Equation(\ref{eqn30}) for \textit{$p_{\phi} = 0$} becomes:

  \begin{equation}\label{eqn31}
\frac{\dot {t}}{\dot{x}} = \frac{\left(-\frac{-A e^{2\beta}e^{2\gamma} + A E^2}{e^{2\beta}(e^{2\gamma})^2}\right)}{\left(\frac{ - A^2(-1 + \frac{E^2}{e^{2\gamma}e^{2\beta}}) + e^{2\gamma}}{e^{2\beta} e^{2\mu} (e^{2\gamma})^2}\right)^{\frac{1}{2}}}.
\end{equation}

   \item Non-Zero Angular Momentum Particles ($p_{\phi} \neq 0$) 
  
  For \textit{$p_{\phi} \neq 0$} equation (\ref{eqn30}) remains same, i.e:
  
 \begin{eqnarray}
\frac{\dot {t}}{\dot{x}} = \frac{\left(-\frac{-A e^{2\beta}e^{2\gamma} - p_(\phi)e^{2\gamma} + E + A E^2}{e^{2\beta}(e^{2\gamma})^2}\right)}{\left(\frac{\frac{2Ap_{\phi}E}{e^{2\beta}} -( \frac{p_{\phi}^2 e^{2\gamma}}{e^{2\beta}} + A^2(-1 + \frac{E^2}{e^{2\gamma}e^{2\beta}}) + \frac{p_{z}^2 e^{2\gamma}}{e^{2\mu}} + e^{2\gamma})}{e^{2\beta} e^{2\mu} (e^{2\gamma})^2}\right)^{\frac{1}{2}}}. \nonumber
\end{eqnarray}
 \end{itemize}

Using the values from equations (\ref{eqn14}) , (\ref{eqn15}), (\ref{eqn16}) , (\ref{eqn17}), (\ref{eqn18}) in (\ref{eqn31} and \ref{eqn30}) , the space-time diagram for radial timelike particles with $p_{\phi} = 0$ and $p_{\phi} = 1$ are obtained.
The numerical values of the constants used here are the same as used in the previous section.

\begin{figure}[h!]
\centerline{\includegraphics[scale=.7]{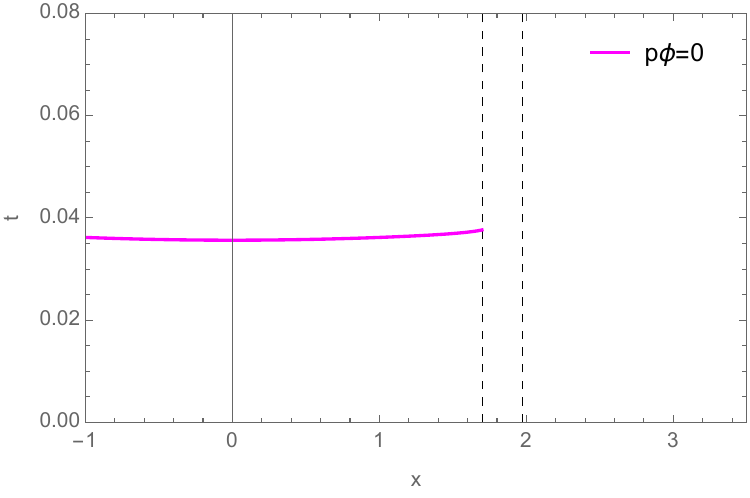}}
	\caption{Space-time diagram for radial time-like geodesic with $p_{\phi} = 0$.
	Here the values used are: $x_{0} = 3.5, \rho_{0}= .4, r_{0} = 1, \chi=.5, m=E_{0}=0, A = 3$. Particles are confined within $x= 1.697$.}	\label{plot4}
\end{figure}

\begin{figure}[h]
	\centerline{\includegraphics[scale=.7]{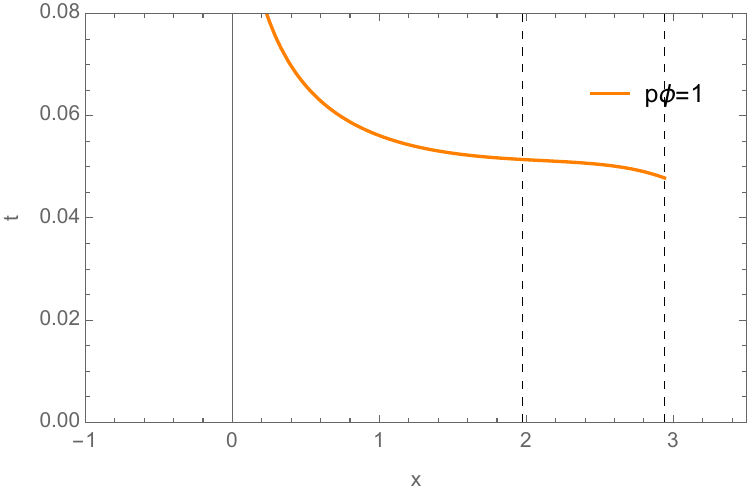}}
	\caption{Space-time diagram for radial time-like geodesic with $p_{\phi} = 1$.
	Here the values used are: $x_{0} = 3.5, \rho_{0}= .4, r_{0} = 1, \chi=.5, m=E_{0}=0, A = 3$. Particles are confined within $x= 2.940$.}
	\label{plot5}
\end{figure}
Interestingly, similar confinement characteristics are seen  in the space-time diagram and also in velocity profile of the radial time-like particles which supports only particles with AM crossing the CTC boundary.

\begin{figure}[h!]
	\centerline{\includegraphics[scale=.7]{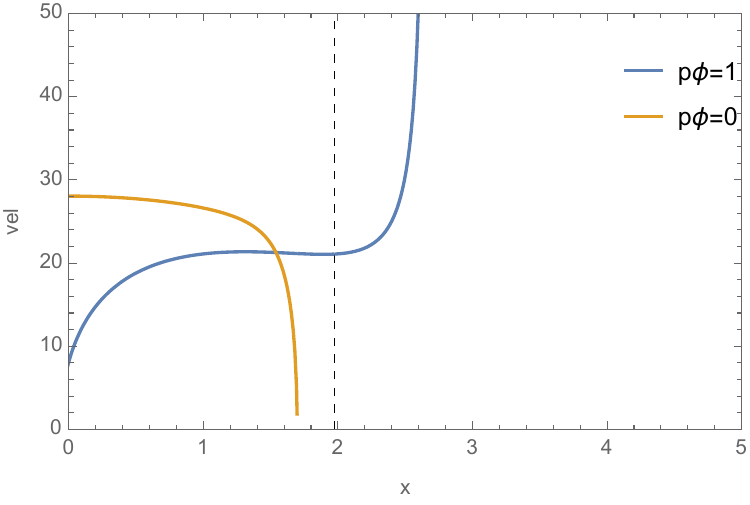}}
	\caption{Velocity profile graph for radial time-like particles with $p_{\phi} = 0$, and $p_{\phi} = 1$. Particles with AM can only cross the CTC boundary denoted by the dashed lines.}
	\label{plot6}
\end{figure}
 
Another effective way to study the dynamics of the motion of the radial time-like particles in the constant Z-plane , we consider the effective potential approach. The motion of the test particles here is considered as the motion of a classical particle in one dimension in the effective potential $V(x)$. The effective potential for the geodesics (considering no motion in the Z- coordinate) is given by \cite{DP,CW}

\begin{equation}\label{Potential}
V(x) = \frac{-b \pm \sqrt{b^2-4ac}}{2a}.
\end{equation}

The above expression can be easily derived from the radial time-like geodesic equation. The values of a, b and c in this case is given by
\begin{eqnarray}\label{values}
\nonumber a &=& -\left(-1 + \frac{E^2}{e^{2\gamma}e^{2\beta}}\right), \\ 
\nonumber b &=& -\left(\frac{2p_{\phi}E}{e^{2\beta}}\right), \\
c &=& -\left(\frac{p_{\phi}^2 e^{2\gamma}}{e^{2\beta}} + \frac{p_{z}^2 e^{2\gamma}}{e^{2\mu}} + e^{2\gamma}\right).
\end{eqnarray}

Using equations (\ref{eqn14}) , (\ref{eqn15}), (\ref{eqn16}) , (\ref{eqn17}), (\ref{eqn18}) for the values of a, b and c one may obtain the effective potential for the radial time-like particles from equation (\ref{Potential}).

Effective Potential is plotted against the radial distance to investigate the dynamics of the time-like particles.

\begin{figure}[h!]
	\centerline{\includegraphics[scale=.7]{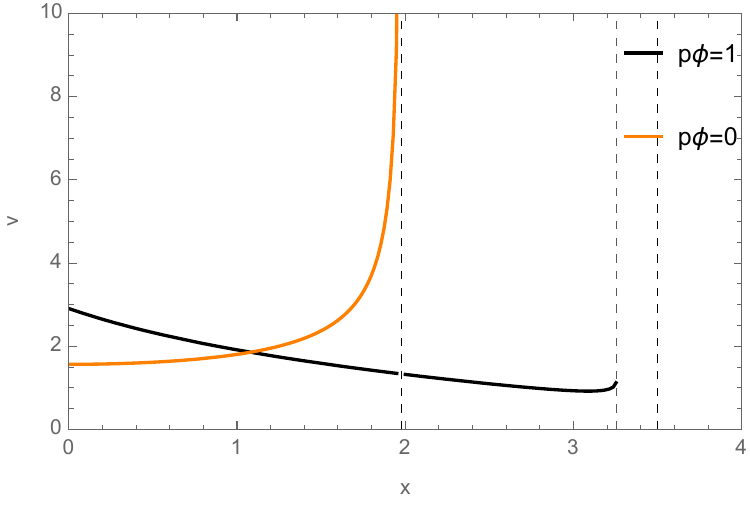}}
	\caption{Effective potential $v/s$ radial distance graph for particles with $p_\phi = 0$ and $p_\phi = 1$. The vertical dashed line (at x=1.974 )represents the boundary of the CTC}
	\label{plot7}
\end{figure}

Figures (\ref{plot4}) and (\ref{plot5}) illustrates the spacetime diagram for particles with Zero AM and particles with Non-zero AM. The particles here are confined within the much less values of ``x" as compared to Photon confinement, as it should be.The Zero AM particles are confined with in the region $x = 1.697$, where as particles having Non-zero AM are confined with in $x = 2.940$. This also can be explained,  as the result of the Total Effective AM  which allows the particle to breach the CTC boundary, and traverse in that region. 
Furthermore, this characteristic can also be explained via the effective potential of the particles. Particles with Zero AM posses an attractive potential, where as particles with AM has an repulsive behaviour of the potential as evident from Figure (\ref{plot7}). 
The potential of the AM less particles acts as a barrier when the particles reaches near its confinement and is pulled back. This can also be observed in the velocity profile figure (\ref{plot6}). The velocity decreases as the radial distance increases, but does not reach zero. The particle may oscillate in that given region.The same potential also prevents the particle from reaching the singularity region as it asymptotes near the confinement region. Velocity Profile of the above said particles for the same values used also supports the above finding. 

The repulsive nature of potential of  particles with AM  can be seen in the figure (\ref{plot7}). The potential also help the particles to cross the boundary and traverse within the CTC region. Initially the velocity of the particle increases as the radial distance increases, helping it to cross over to the causality violating region but ceases to increases after a certain time. This shows that even if the potential is repulsive it restricts the particle to have any motion near the singularity. 

Moreover , Closed Timelike Curves  generally occur when the other coordinates(t,r,z) except the $\phi$ coordinate are considered constant \cite{Visser,lobo}. This allows the particle to only have equatorial motion. Now using this equation our metric reduces to:
\begin{equation}\label{eqn34}
ds^2= \left[-\frac{E^2}{e^{2\gamma(x)}}+e^{2\beta(x)}\right]d\phi^2 .
\end{equation}

Using the Euler-Lagrange equation one may obtain the canonical  momentum of the $\phi$ coordinate, which is:
\begin{equation}\label{eqn35}
\dot \phi = \frac{p_\phi}{\left[-\frac{E^2}{e^{2\gamma(x)}}+e^{2\beta(x)}\right]} .
\end{equation}
From equation (\ref{eqn35}) it is clear that in CTC particle with zero angular momentum cannot exist as, if $p_\phi$ becomes zero , $\dot \phi$ becomes zero which in turn makes `$\phi$' a constant. This violates the basic condition for CTC to form as all other coordinates except $\phi$ needs to be constant. Hence zero AM particles are forbidden in the causality violating region.

\section{Conclusion}
The results obtained in this article helps in better understanding of particle motion along with its confinements and its characteristic near causality violating regions. It also provides insights on traversability through Cylindrically Symmetric Wormholes admitting to Closed TImelike Curves.

In this article we made a detailed analysis on the motion of test particles in the Cylindrical Wormhole spacetime. We also studied the characteristics  of test particle motion in the vicinity of CTC and their confinements. The studies made in this paper show that angular momentum plays an important part in the confinements of the particles, and their study is of great importance in analysing the particle motion. The present work is novel in the sense that the geodesic motion of particles in the Cylindrical Wormhole spacetime along with their dependence on the angular momentum has not been studied. In section (\ref{III}) the spacetime dragging term obtained from the geodesic equation made it clear that particles without AM also co-rotates with the spacetime. As a result analysis of confinements of AM less particles along with particles with AM is also of great interest. 

In this study it is shown in details that the particles without AM are confined outside the CTC boundary where as particles with AM can penetrate the CTC boundary to traverse within the closed orbit. This happens due to the Total Effective AM which increases the particle's effective acceleration, which in turn increases the velocity of the particle allowing it to move past the barrier where causality violating region occurs. By solving the  geodesic equation , the spacetime diagram for radial null and radial time-like particles are obtained along with their respective confinements. Its clearly shown in the figures \eqref{plot1}, \eqref{plot2}, \eqref{plot4}, \eqref{plot5} that only non-zero AM particles can traverse within the CTC.

We also studied the effective potential of the particle which is another way to analyse the dynamics of the motion of test particles.The analysis of the effective potential helps study the  characteristics of the particles. It is shown that the potential for a AM less particle is attractive in nature forbidding it to cross the CTC confinement. It also restricts the particle from travelling to the singularity region. The particle oscillates between the region of its confinement. This claim is supported by the velocity profile as it indicates, decrease in velocity with increase in radial distance but velocity of the AM less particles never becomes zero. In contrast the potential of non-zero AM particles is repulsive in nature allowing it to cross the confinement of the CTC, but forbidding it to reach the singularity region. The velocity of the particle supports this as the velocity increases with the  increase in radial distance, but asymptotes before the singularity. Further, it is also shown in general that the said spacetime admitting to CTC cannot have particles without AM to traverse within the curves. Considering particles with zero angular momentum violates the conditions put forward for CTC to form, hence, forbidding them to be present in the region. 

The present work shows an explicit example in which test particles can move on CTC and co-rotate with the wormhole instead of counter-rotating though in the literature most of the examples are actually counter-rotating (for example Van Stockum spacetime \cite{AD}). This implies that CTC is not the result of frame-dragging (see ref. \cite{Andreka:2007ai}) and this is a generic phenomenon which was challenged in \cite{Duan:2021pci}. Therefore, our model supports the result in ref. \cite{Andreka:2007ai} that frame dragging is not needed in CTC formation.

\section*{Acknowledgement}
The authors are thankful to the Inter University
Centre for Astronomy and Astrophysics (IUCAA), Pune (India) for their hospitality, as a part of this work was done during a visit there. S.C. thanks FIST program of DST, Department of Mathematics, JU (SR/FST/MS-II/2021/101(C)).


\begin{thebibliography}{50}

	\bibitem{ER}
 A. Einstein, N. Rosen, Phys. Rev. 
 \textbf{48}, 73 (1935)
 
	
	\bibitem{MT}
 M.S. Morris, K.S. Thorne, Am. J. Phys. 
 \textbf{56}, 395 (1988)

 \bibitem{Visser}
 Matt Visser. Lorentzian wormholes: from Einstein to Hawking. 
 \textbf{1995}.

\bibitem{CNHC}
D.Roy, A.Dutta, S.Chakraborty, EPL \textbf{140} 19002 (2022)

\bibitem{hocheberg_visser}
 David Hochberg and Matt Visser, Phys. Rev. D \textbf{58}, 044021 (1998)

 \bibitem{Hayward}
 S.A. Hayward, IJMPD \textbf{08}, 373 (1999)
	
 \bibitem{bronnikov1}
K.A. Bronnikov, Phil. Trans. R.Soc.A \textbf{380}:20210176
\bibitem{bronnikov2}
K.A. Bronnikov and V.G Krechet, IJMPA \textbf{31} (2016)
\bibitem{bronnikov3}
K.A. Bronnikov and Lemos Phys. Rev. D \textbf{79}, 104019, (2009)
\bibitem{bronnikov4}
K.A. Bronnikov and V.G Krechet, Phys. Rev. D \textbf{99}, 084051, (2019)

\bibitem{Forghani}
S.D. Forghani \it{et.al}, JCAP10(2019)067

\bibitem{Eiroa}
Eiroa and Simeone , Phys. Rev.D \textbf{91}, 064005 (2015) 	

\bibitem{Cataldo}
Mauricio Cataldo, \textit{et.al}, Eur. Phys. J. C \textbf{77}(11), 748, 2017	

\bibitem{Olmo}
Gonzalo J. Olmo,\textit{et.al}, Phys. Rev. D \textbf{92}, 044047, 2015

\bibitem{CC}
Chandrachur Chakraborty and Parthapratim Pradhan , JCAP03(2017)035

\bibitem{Ellis}
Homer G. Ellis, J. Math. Phys. 14, 104 (1973)


\bibitem{lobo}
F. S. N. Lobo,
\textit{arXiv:1008.1127} (2010)

\bibitem{FW}
Felix Willenborg, \textit{et.al}, Phys. Rev. D \textbf{97}, 124002 , (2018)

\bibitem{DP}
D. Pugliese, H. Quevedo, and R. Ruffini, Phys. Rev. D
\textbf{88}, 024042 , (2013)

\bibitem{CW}
C.W. Misner, K.S. Thorne, and J.A. Wheeler, Gravitation, W.H. Freeman (1973)

\bibitem{AD}
A. Dutta, D. Roy, and S. Chakraborty, Int. J. Mod. Phys. D 31(13), 2250096 (2022)

\bibitem{Andreka:2007ai}
H.~Andreka, I.~Nemeti and C.~Wuthrich,
Gen. Rel. Grav. \textbf{40}, 1809-1823 (2008)
doi:10.1007/s10714-007-0577-1
[arXiv:0708.2324 [gr-qc]].

\bibitem{Duan:2021pci}
Y.~Duan, F.~Liu, Y.~Wang and Y.~C.~Ong,
Universe \textbf{8}, no.1, 28 (2022)
doi:10.3390/universe8010028
[arXiv:2107.08844 [gr-qc]].

\end{thebibliography}
\end{document}